# A Pragmatic Approach to Sovereignty on Mars

Sara Bruhns and Jacob Haqq-Misra

Blue Marble Space Institute of Science
1001 4th Ave, Suite 3201, Seattle, Washington 98154, United States
Email: jacob@bmsis.org

**Abstract**

Rising interest in Mars colonization from both private and public sectors necessitates a renewed discussion about sovereignty in space. The non-appropriation principle of the Outer Space Treaty currently prohibits any sovereign claims to celestial bodies, but it remains unclear how this principle should be applied to the peaceful colonization of Mars. Here we develop a pragmatic approach to guide the settlement of Mars, which is based upon a "bounded first possession" model with mandatory planetary parks. Scientists, experts, and leaders will establish planetary park locations and regulations through worldwide community solicitation in order to protect sites of scientific, aesthetic, historical, cultural, environmental, spiritual value. Colonization parties may occupy limited plots of martian land and may claim exclusive economic rights within this zone, while still refraining from any claims to sovereignty. All colonists remain under the legal jurisdiction of their host nation, with conflicts to be resolved diplomatically or through a temporary tribunal system composed of representatives from other Mars colonies. We also propose the formation of a Mars Secretariat as an administrative body with limited power to facilitate communication among parties. Our model for Mars colonization remains consistent with the Outer Space Treaty, but we also recommend revisiting or amending the non-appropriation and province of mankind principles to resolve the ambiguity of how nations, corporations, and individuals may utilize the resources of space.

**1. Introduction**
As national space agencies and private aerospace corporations develop the technology to send larger groups of humans into space for longer periods of time, Mars colonization faces an increasingly probable future. A more significant human presence in space will test the resilience of the Outer Space Treaty (OST), which entered into force in 1967, just before humans landed on the Moon. Thus far, the OST has been successful in maintaining peaceful international relations in space. The OST decrees that "the exploration and use of outer space shall be carried out for the benefit and in the interests of all countries and shall be the province of all mankind" to ensure that space is "free for exploration and use by all States." Specifically, the OST explicitly states that "outer space is not subject to national appropriation by claim of sovereignty, by means of use or occupation, or by any other means," a requirement known as the *non-appropriation principle* of the OST [1]. Any celestial bodies must be used for exclusively peaceful purposes, and no State may launch weapons into space. States are also responsible for all national activities, whether governmental or otherwise, liable for any damage caused by their space objects, and must not contaminate celestial bodies. The OST has been upheld thus far and provides the primary legal requirements for any further exploration, including colonization, of Mars [2].

Current colonization proposals from organizations such as SpaceX and Mars One may be inconsistent with the principles of the OST. Commercial interests may drive private corporations toward the profitable use of space resources [2], despite objections that such actions would violate the non-appropriation principle [1]. The current international legal environment mandates that each nation is responsible for any spacecraft launched from their territory, and any corporation venturing into space would likely continue to fall under the jurisdiction of its nation of incorporation [1,2]. However, the possibility remains that a multinational corporation could establish operations that circumvent the OST in order to gain priority access to space resources. Even if a corporate actor falls within jurisdiction of the OST, enforcement of the OST may prove difficult for the first few colonies that develop on Mars. In the event that any such corporations land on Mars and begin claiming resources, such actions will be much more difficult to modify if guidelines on legal colonization and utilization of space are never established beforehand.

Ehrenfreund et al. [3] recognize the importance of approaching Mars settlement with an emphasis on working to balance the diverse perspectives among scientific, governmental, and private interests. Their approach recommends gathering information on all potential stakeholders, cooperatively developing a timeline and drafting agreements, and legally establishing a system that remains flexible to accommodate changing and new interests as they arise. The goal is to adhere to the OST and balance science exploration and use by government and industry while minimizing any harmful environmental impacts of colonization. In particular, Eherenfreund et al. [3] discuss possible models for space exploration by drawing upon the Antarctic Treaty System (ATS) and the UN Convention on the Law of the Seas (UNCLOS) as examples of successful sharing of international resources. This attention to the diverse interests of stakeholders in the martian landscape represents a compromise approach to the issue, where inclusivity based upon contemporary international treaties seeks to forestall future conflict.

The commercial potential of celestial resources is a powerful incentive, and Collins [4] argues that corporations will inevitably colonize with sufficient economic motivation. However, the OST suggests a common property approach through the *province of mankind principle*[1], which limits the reward incentive to offset the cost and risk of pioneering colonists. This may discourage investment and productive use of planetary resources, which ultimately could be harmful or restrictive to the future of humanity. Collins [4] proposes that private property claims be permitted in order to encourage the efficient preparation of Mars for human habitation, either through modification of the OST or by other actions of international governing bodies such as the United Nations. In order to reward as many pioneers as possible, Collins [4] suggests a model of "bounded first possession by landfall" (hereafter, "Bounded Possession") of limited plots of land, controlled by private bargaining and litigation, as one possible option. This would allow national space agencies or private corporations to make small exclusive claims to martian land, allowing for the development of settlements and industry but still leaving plenty of surface area for future colonists. Such a policy would require modification of the non-appropriation principle of the OST in order to legally permit sovereign claims to space resources [1]. Another option proposed by Collins [4] is a "Mars Tax," collected in some way from the use of Mars, which would be distributed to all nations in the spirit of the province of mankind principle of the OST. A Mars Tax would allow Mars to be used by those for whom it is most profitable, but the portion of these profits paid as a

---

[1] We use the phrase "province of mankind principle" to remain consistent with the phrasing of the Outer Space Treaty, although "humankind" would be a more preferable term to include in any revisions of the OST.

Mars Tax would thus provide direct benefits to all nations, not just those with spacefaring technology. The use of a Bounded Possession system or a Mars Tax, either separately or in tandem, would require amendment to or replacement of the OST, but solutions such as these may provide the strongest motivation for the productive development of space.

Scientific objectives have formed the basis for nearly all space exploration up to now, and the collective interests of scientists will also shape the future direction of any policies that govern the exploration of Mars. Scientists hold a broad range of views on colonization [5]. Some are enthusiastic about a more prolonged and personal experience with the planet [6,7], while others are troubled over the inevitable contamination [8]. Cockell and Horneck [6,7] suggest a "Planetary Park" system that preserves land for scientific, cultural, and aesthetic purposes. These parks would represent a diverse portion of Mars' terrain and would be regulated to minimize human contamination. These parks may also fulfill the OST province of mankind principle, perhaps opening the possibility to free industries and nations from any "common property" stipulations and allow the development of property ownership elsewhere on Mars. Cockell and Horneck put forth a list of suggested park locations that represents the most scientifically valuable land on the planet. While a Planetary Park system focuses on protecting the scientific and aesthetic value of Mars, this solution may also appeal to corporate interests as a compromise between protecting environmental interests and allowing private use of resources.

We propose a model for Mars exploration and settlement that combines many aspects of these approaches. We consider the successes and shortcomings of the ATS and UNCLOS as models for cooperative sovereignty in space [3], we discuss a system of exclusive economic claims under a Bounded Possession policy [2,4], and we incorporate a Planetary Park system designated though assessment by the worldwide community of scientists, experts, and leaders [6,7]. Regardless of our opinions on the ideals of land ethics, we propose our model on the pragmatic assumption that private space industry colonization missions will be carried out as intended. We assume that colonization is inevitable, and that restricting sovereignty may pose as many as or more problems than allowing it. We present our "Bounded Possession with Planetary Parks" system as one possible solution toward resolving the issue of sovereignty on Mars. Our model allows the global community to establish a Planetary Park system prior to humans landing on Mars, which colonists must respect. Parties landing on Mars will be allowed to claim a limited amount of land, and only as much as they can reasonably use. This provision is reminiscent of ideas described by John Locke in his *Second Treatise on Government*, where the valuation of land is proportional to the amount of human labor invested in it. The colonies will be self-governing and independent, but must obey the guidelines of the planetary parks, follow procedures for making an exclusive economic claim, and respect the boundaries of other colonies. Our model also includes a "Mars Secretariat", an exclusively administrative body, to facilitate communication between the colonies and act as a mediator in the event of conflict between parties. Any conflicts that cannot be resolved between the parties involved will be addressed by a temporary tribunal led by representatives from the other colonies and facilitated by the Mars Secretariat. This model of Bounded Possession with Planetary Parks allows for colonization to proceed in a cooperative manner while mitigating conflict escalation.

## 2. Cooperative Sovereignty on Earth Today
The Outer Space Treaty prohibits nations from declaring sovereign claims on any celestial bodies. One motivating factor for this provision, drafted in the midst of the Cold War, is that sovereign claims in space could foster militarization and escalate conflict between nations [9,10]. While disputes over sovereignty are plentiful in history, there are also several examples of situations where sovereign nations work together cooperatively to achieve mutual goals, instead of generating conflict about borders and control. The concept of "cooperative sovereignty" even suggests that cooperation among states is a critical element of the notion of sovereignty itself in an increasingly interdependent world [11]. The ATS and UNCLOS both provide contemporary examples of models based upon principles of cooperative sovereignty.

The Antarctic Treaty System provides a constructive model of nations cooperatively managing land for solely scientific purposes. Nations party to this treaty have surrendered their economic and sovereign interests in favor of preserving natural, unspoiled land and furthering science for the benefit of all nations. Drafted in 1959 after tensions about land claims escalated near conflict, the ATS represents an unprecedented level of international cooperation and an unorthodox attitude about land claims. It prohibits all military activities, encourages freedom of science and sharing of information, requires peaceful settlements of disputes, bans non-science mining, and prohibits new land claims. The original signatories and subsequent nations who signed and are recognized by the others constitute the parties of the ATS, who consult together to pass treaty amendments. A central body known as the Antarctic Treaty Secretariat performs administrative tasks that include management support of the annual meeting of the Parties. The focus of the treaty is clear: peaceful governance of the exclusively scientific activities on Antarctica for the mutual benefit of all.

The Law of the Seas provides an example of another way to manage shared space. More precisely known as The United Nations Convention on the Law of the Sea (UNCLOS), the first and second versions were succeeded by the third version in 1982. This treaty was influenced by the UN Trade and Development Conference and calls for fairer terms for developing nations in the form of technology and wealth transfers from richer to poorer countries and pollution control. The USA has not ratified the latest version of the Law, objecting to its redistributionist tendencies. However, the United States recognizes the territorial and regulatory provisions of UNCLOS as customary international law. Each nation's sovereign territorial waters extend to a maximum of 12 nautical miles beyond its coast. Foreign vessels are granted innocent passage through this area, but must refrain from engaging in military, pollution, fishing, or science activities. Each nation may also establish an exclusive economic zone out to 200 nautical miles off its coast. In this zone, the country can exploit and regulate fisheries, construct islands and other installations, regulate science on foreign vessels, or conduct other economic activities. Foreign vessels may still move freely through this space. Each nation has exclusive rights to seabed resources up to 200 miles from its shore or to the continental margin, whichever is further, with a 350 mile limit. Where any of these borders overlap between nations, boundaries must be drawn by mutual agreement. Beyond these boundaries lie the high seas, which are open to all nations.

These treaties have succeeded in deferring the interests of individual nations in favor of the common benefit of all. The cooperative models of the ATS and UNCLOS thus provide a framework for managing the shared use of space resources [3]. However, we also identify two elements from the history of these treaties —the lack of strong central authority, and objections to equitable sharing—that will likely affect any approach to sovereignty on Mars.

*2.1 Lack of Strong Central Authority*
The ATS has been successful in focusing attention solely on scientific pursuits and suspending any sovereignty issues, but many of the Parties to the ATS were unwilling to permit the emergence of a powerful separate central authority. A central body known as the Secretariat was rejected by many of the pioneering Parties (Argentina, Chile, Australia, and France) in all forms except for a minor administrative role. Most Parties disliked the idea of any significant centralized governmental authority that would interfere in their affairs, preferring to work amongst themselves to resolve issues. The Secretariat that presides over Antarctica today has little authority and amounts to a supportive administrative body.

For similar reasons, many nations declined to sign the Moon Treaty of 1979, objecting to the establishment of "an international regime, including appropriate procedures, to govern the exploitation of the natural resources of the Moon as such exploitation is about to become feasible." The formation of a powerful centralized body is one of the primary reasons that the Moon Treaty has not been ratified by any space-faring nations [10]. Visions of central entities such as a World Space Agency [12,13] or an Interplanetary Authority [14] for future space exploration may be met with similar resistance, while centralized authority in the space environment could also foster abuses of power that curtail civil liberties [15,16].

The unpopularity of new central authority is an important lesson to keep in mind for any successful model that applies to Mars governance, as any strong central authority is likely to be rejected by parties that have nothing to gain by giving up their autonomy. We will therefore assume that any successful model for sovereignty on Mars cannot include a strong centralized authority.

*2.2 Objections to Equitable Sharing*
Another objection in international ownership treaties is the concept of equitable sharing. The UNCLOS includes a commercial mining section that deems minerals in the seabed and ocean floor of the high seas as "the common heritage of mankind" and places them under the jurisdiction of the International Seabed Authority (ISA). Commercial exploration and mining have not yet occurred but would be regulated by the ISA in the spirit of assisting developing countries. It was largely this section of UNCLOS III that the United States and other developed countries objected to. If commercial mining of the high seas were to begin, the ISA would establish a global mining enterprise, which would be at least equal in size and scope to any private or state enterprises. Fees and royalties from the private and state ventures as well as profits from the global sector would be distributed to developing countries. Private and state miners would also be strongly encouraged to sell their technology and expertise to the developing countries. The US and other industrial countries objected to this redistributionist requirement as unfavorable to economic and security interests. These countries convened to negotiate amendments to UNCLOS that espoused more free-market principles. An agreement was reached and the US signed but did not ratify the Treaty. A decline in demand for minerals and the fall of Communism made the seabed issue less relevant and several calls for ratification within the US government have been made and positively met.

The Moon Treaty also includes similar language providing that "the Moon and its natural resources are the common heritage of mankind," which constitutes a legal right of all nations of the world to the resources of space. This equitable sharing principle is unpopular among space-faring nations, particularly those that have invested heavily in the technology required to enable space exploration.

These objections to equitable sharing also provide lessons for sovereignty in space. All space-faring nations have signed the OST and must abide by the province of mankind principle, but all have rejected the common benefit of mankind principle of the Moon Treaty. The current legal environment makes space free for use by all but makes no requirements for redistribution. We will therefore assume that any successful model for sovereignty on Mars cannot require equitable sharing.

## 3. Challenges to Cooperative Models of Sovereignty

The above section describes existing treaties for international ownership that have broadly succeeded. The ATS provides a system that functions well for the scientific exploration of Antarctica, and the UNCLOS allows for peaceful use of the seas for commerce and science. In the subsections below, we discuss some challenges to this model of sovereignty, which can provide insight into developing an improved approach to sovereignty in space.

*3.1 Humane Society International v Kyodo Senpaku Kaisha Ltd*

The question of sovereignty in Antarctica has been a largely unresolved issue. Sovereign claims of Antarctic territory were made in the early-mid 1900s by the UK, Chile, France, Norway, and Argentina based on the traditional grounds of discovery, occupation, and geographic proximity. The international community rejected and ignored these claims, establishing their own ventures on the continent. Tensions rose and the unorthodox treaty was surprisingly agreed upon. The ATS prohibits any new land claims, and states that the treaty in no way supersedes any land claims previous to the drafting. In effect, sovereignty claims in Antarctica are honored superficially, but practically they are essentially ignored. Theoretically, the exclusively scientific and cooperative sharing of information embodied in the spirit of the ATS should make sovereignty a non-issue, and each nation traditionally requires its nationals to follow their own laws, as if they were living in their home nation. But when people of other nations come into a land claimant's area and violate the laws of the claimant's nation, a dilemma is introduced. In order to preserve the land claim, this nation would want to penalize the lawbreaker and exercise their sovereignty. But the ATS encourages a sharing of resources which has been successful at minimizing conflict. To enforce a national law, this nation would now be pressing the ambiguity of the sovereignty issue of the ATS, thereby jeopardizing the success of the treaty and its own interests.

A recent example of this is the 2004-2008 *Humane Society International Inc. v Kyodo Senpaku Kaisha Ltd.* court case [17,18]. Japanese whalers began killing Minke whales in Australia's part of Antartica, violating Australia's Environment Protection and Biodiversity Conservation Act. Kyodo had been granted a special whaling permit by the Japanese government which allowed them to circumvent commercial whaling laws. Japan argued that they did not recognize Australia's land claim, that they were conducting scientific whaling, and that they were acting in accordance with the Antarctic Treaty System. However, the region of Antarctica under dispute was also part of the Australian Whale Sanctuary, and by their law citizens of any country were prohibited from whaling in this protected area. Kyodo was therefore in violation of Australian law. The Humane Society International Inc. sought enforcement based on these Australian laws. Initially, Australian courts rejected the case based on the inability of Australia to enforce any ruling of the court (due to the ambiguity of sovereign claims under the ATS) and in hope of avoiding international conflict with Japan. The court eventually accepted the case despite these concerns and ruled that Japan had violated Australian law.

This ruling has not been enforced[2] but represents an important issue with the ATS. If all Parties work cooperatively and focus on cooperative goals, then the ATS functions adequately and no sovereign claims need to be asserted. But if one or more nations enter into conflict, then the ATS will begin to show its deficiencies. Nations that have made sovereign claims will seek to defend their claims, while nations without existing claims may object to honoring any claims to sovereignty while the ATS remains in effect. As long as no country encroaches on another there are no issues, but as soon as conflict between Parties occurs in Antarctica, the ATS will break down.

---

[2] A separate case heard by the International Court of Justice titled "Whaling in the Antarctic" (Australia v. Japan: New Zealand intervening) ordered a temporary halt to Japan's Antarctic whaling program on 31 March 2014 with a ruling that the whaling expeditions lack scientific merit.

The *Humane Society v Kyodo* case is a pertinent example of a challenge to this cooperative model that could call into question any ability of the ATS to resolve disputes over sovereignty [19]. Even with the goal of maintaining an open environment of scientific cooperation, the ambiguity of sovereign claims under the ATS leaves room for unresolved conflict. We will therefore require that any successful model for sovereignty on Mars includes an explicit procedure for conflict resolution among parties.

*3.2 Moratorium on Commercial Activities in Antarctica*
Antarctica has been almost exclusively used for scientific purposes thus far, and information has generally been shared freely, in the spirit of the treaty. Part of the reason this cooperation is possible is that the Protocol on Environmental Protection to the Antarctic Treaty establishes what essentially amounts to a 50 year moratorium on non-scientific mining. Until the year 2048, no non-scientific mining may occur in Antarctica unless there is unanimous agreement by all parties and a binding legal regime on Antarctic mineral resource activities is in force. After this time, a 3/4 majority of the Consultative Parties may overturn the prohibition, but until then non-science mining remains illegal. The cooperation of signatory parties has therefore been mostly untested, and commercial interests may ultimately lead to the end of the ATS in 2048 with the potential initiation of non-scientific mining. Mineral resources in Antarctica may prove difficult to extract and obtain a return on investment in the near future [20], but non-scientific motivations may ultimately change the cooperative political landscape of Antarctica today. Once land claims in Antarctica become commercially valuable, the ATS will have much more difficulty in maintaining its ambiguous stance toward national sovereignty.

The success of the ATS offers a cooperative model for peaceful exploration, and perhaps even the eventual initiation of mining will still allow the ATS to uphold the original spirit of scientific priority, preservation of valuable natural landscape and ecosystems, and cooperation amongst all nations. Commercial interests in space resources are growing, and a moratorium that would limit these activities may ultimately delay or hinder the human exploration of space. We will therefore assume that any successful model for sovereignty on Mars cannot require a moratorium on commercial activities.

*3.3 The Great Pacific Garbage Patch*
The UNCLOS handles border definitions and transit regulations with success, but the common property approach toward management of the high seas has created a significant problem of the commons [21]. The Great Pacific Garbage Patch is a giant gyre of debris suspended by currents in the northern Pacific Ocean [22]. The patch is composed mostly of plastics and organic pollutants [23] and estimates of its magnitude range from the size of Texas to the size of the continental US. Other similar floating patches of debris are found in the southern Pacific [24] and other regions of the world's oceans. Some researchers have begun to investigate methods for reducing the amount of plastic debris in these expansive patches [25], but the UNCLOS provides no enforcement or incentives to address this problem.

The persistence of these oceanic debris patches highlights the inability of the UNCLOS model to effectively manage shared resources. This type of problem of the commons in the world's oceans also serves as an example for the management of resources in space. Although the physics of transport by oceanic currents is unique to Earth, we will assume that any successful model for sovereignty on Mars must include a procedure for addressing problems of the commons.

**4. A Pragmatic Model for Sovereignty on Mars**
Given the numerous initiatives devoted to Mars exploration and colonization in both the public and private sectors, we assume that colonization of Mars will eventually occur, through either or both corporate and government efforts [2,4]. Economic motivations for developing the resources of space are growing, but humanity's spirit of adventure also cannot be discounted as a significant motivation for colonization. History has shown that humans have explored and exploited nearly all available land on Earth in less than 10,000 years, and we expect a much faster trajectory for planetary colonization due to the rapid pace of technological advances [2,26]. In an attempt to accept the inevitable and in order to staunch the potential harmful effects of human colonization of another planet, we propose a model for colonization that will preserve large swaths of Mars for scientific, aesthetic, historical, cultural, environmental, spiritual, and a variety of other purposes as well as allow human settlement with economic and commercial goals to occur. We suggest that, regardless of any particular choice of normative ethical position on the use or valuation of land and resources, it is unrealistic to expect humanity to forgo colonization when it is possible, profitable, and potentially a source of useful resources. With this sentiment as our basis, we propose a model based upon bounded first possession by landfall with large-scale planetary parks for the settlement of Mars. This will allow for the inevitable colonization but avoid competing sovereignty claims as well as preserve critical regions for both science and overall preservation. We suggest this Bounded Possession with Planetary Parks model as a balanced compromise between scientific exclusivity, national sovereignty, and corporate control of Mars.

*4.1 Planetary Parks*
According to our model, the international community would designate a planetary park system on Mars that would absolutely exclude any colonization or unregulated passage or usage within the park boundaries. Cockell and Horneck [6] suggest a series of guidelines as well as suggestions for the first seven parks. Their guidelines are 1) no spacecraft parts left in the parks 2) no landing of unmanned spacecraft 3) no waste 4) foot or surface vehicle access only in predesignated areas and 5) suits must be externally sterilized. Cockell and Horneck [6] suggest these seven initial parks: 1) polar 2) Olympus 3) desert 4) historical (the sites of rovers) 5) Marineris 6) Southern and 7) Hellas. These are excellent sites, but we recommend the selection to be entrusted to the global community. We suggest that the global community of scientists, experts, and leaders decides through consensus on locations of these initial parks and on the upkeep of the current parks and opinions about new park sites. By following a model similar to NASA's solicitation of community input through decadal surveys by the National Research Council that recommends the next generation of space missions, we suggest that the planetary parks system should be regularly reviewed and updated in order to protect sites of scientific or aesthetic value. NASA's decadal survey process is a method for obtaining community input to identify key research questions and identify the next generation of space missions. This consensus-based model has shown success among earth and space scientists by allowing a wide range of perspectives to be represented, and the model could be applied toward a more interdisciplinary group of experts and leaders to determine the allocation of planetary parks on Mars. Such a model risks isolating minority groups in favor of the consensus of the majority, but a global consensus would at least be more defensible against any commercial interests that seek to infringe upon planetary park boundaries.

The international community, coordinated through a body such as the UN Committee on the Peaceful Uses of Outer Space (COPUOS) or the Committee on Space Research (COSPAR), would also be responsible for establishing the rules of using the parks, and whether some parks will be reserved for solely preservationist purposes while others are open to scientific exploration. The guidelines proposed by Cockell and Horneck [6] provide a robust place to begin the development of an optimal and evolving set of rules by synthesizing the opinions of an international and interdisciplinary group of leaders and experts representing a diverse array of values; however, much more work in ethics and policy is needed in order to adequately determine how to demarcate planetary parks on Mars in a manner that is amenable to all nations. In this way, we hope to establish internationally recognized park locations and the rules that are needed to maintain them and preserve Mars for diverse interests, from scientific to cultural. We also hope that the comprehensive approval of the global community will mitigate the need for further legal action to create these parks and minimize any conflict. Presenting a unified front of diverse international interests will provide a compelling reasoning behind reserving large swaths of land on Mars for non-commercial use.

*4.2 Exclusive Economic Zones*
Once the planetary park system is formed, settlement can commence. Whenever a colonization mission lands, they may occupy a limited plot of land based on what is reasonable for productive use, perhaps an area bounded by a 100 km radius at first. It may be best for this number to be established before colonization commences in order to mitigate future disputes and conflicts. This initial colony size may subsequently be subject to expansion based on need and mutual colony agreement. In order to remain in accordance with the OST, the colonists will land, occupy, and establish governance on the land, but will not make any formal sovereign claims. This exclusive economic zone governed by the colony will be centered on the station, where the host nation's laws apply. Outside of the station, the only claim the colony will have is an economic claim to the resources within their occupied land. They will have exclusive rights to all resources within their radius, but no other influence. Other settlers may land and even occupy land within another colony, but they will have no economic rights to access the resources.

Following this model, one colony may land and establish its zone, while another party may choose to subsequently land in that zone and occupy it for some time. Perhaps this allowance of exclusive economic zones would even provide benefits for the expansion of colonization: new colonists could land and remain close to existing colonies until they can assess a location for a more permanent colony. But it will be in the second colony's best interests to develop other land as soon as possible, at least if they desire to make exclusive economic claims of their own. The host colony of this exclusive economic zone may mine and conduct scientific research on this land, but they must keep it reasonably clear of human detritus and they have no control over other countries activities there unless it pertains to resources.

Within each martian station, the host nation's laws apply to all occupants, whether visitor or host, while outside of the stations, each settler must abide by their own nation's laws. This precedence for national jurisdiction is analogous to requirements of the ATS as well as contemporary interpretations of space law applied to the International Space Station (ISS). The right of peaceful passage through exclusive economic zones is also patterned after UNCLOS, which also restricts the degree of control that a colony can exert over the zone in which it operates.

Though the point where overpopulation on Mars reaches Earth standards is inconceivably far in the future, in order to avoid a space and resource crisis similar to Earth's we propose absolute limits on the fraction of Mars that any one entity can claim and that humans as a whole can claim. We cannot make premature predictions about how this fractional limit should be set, but such a decision can be made once Mars colonization is underway. Through this pragmatic model, we aim to preserve some semblance of the natural Mars environment as well as avoid any one entity controlling too much land and power in Mars politics.

*4.3 Mars Secretariat*
Our model includes no powerful and centralized Mars authority. Each colony will develop its own exclusive governance within the requirements of existing space law and will retain its own authority in resolving conflicts of the planet. The authority to make and amend communal Mars agreements and negotiations will be the sole prerogative of the colonies currently occupying Mars. Previous efforts to establish central authority in similar circumstances illustrate a general resistance to strong, external power. The Moon Treaty has still not been signed by any space-faring nation mainly because of its provision for a strong central authority. The Secretariat of the Antarctic Treaty System has likewise been rejected in any form except its current administrative position. We expect that any countries or corporations that possess the capital, technology, and initiative to colonize Mars will also resist being subject to an outside power with significant strength.

To this end, we propose not a strong central power, but a weaker administrative body similar to the Secretariat of the ATS. This body would be responsible for facilitating communication between the colonies, mediating conflict resolution, and serving as a repository for legal documents. The Secretariat would be established strictly in a subservient role to the colonies of Mars, with the purpose of preserving the interests of these colonies as much as possible. To support this goal, and in order to pay for this administrative body, we suggest the colonies alone contribute either funds or members to support the administration of the Secretariat. This provides a check on any Earth-based institutions gaining significant power or influence on Mars. This organization of the Secretariat would not threaten the power of any of the colonies or their main countries, and it would also keep the occurrence of meaningless conflicts and problems of the commons low by facilitating communication and clarifying policy.

*4.4 Conflict Resolution*
When a corporate, government, or other body establishes a colony, the settlers of the colony must abide by the laws of the country they represent as well as the more general laws of Mars, such as the land claim and exclusive economic zone laws. When entities adhere to this requirement to establish their own exclusive economic zones and their own governing laws, then this will tend to minimize conflict. However, when the specific laws of each colony conflict, as with the *Humane Society v Kyodo* case, our model includes an explicit conflict resolution system to avoid the ambiguity of conflict under the ATS model.

As a first step, and similar to the ATS, parties are encouraged to resolve disputes peacefully themselves using diplomacy. If this fails, a temporary tribunal will be formed composed of representatives from each Mars colony that is not involved in the case. The host country or corporation of each colony will be given the authority to nominate these representatives. If all colonies were to be involved in a case (as is possible, especially in the early stages when there are few colonies, as well as when planet-wide issues are being discussed) representatives would have to be found elsewhere. Perhaps the nations of Earth that are pursuing, but have not yet achieved, colonization would be the best candidates for composing the tribunal. These nations would have interests that most closely correlate to those of the existing Mars colonies and would therefore be qualified to contribute to an informed tribunal process. The Mars Secretariat will provide administrative support for coordinating these proceedings but will remain impartial and uninvolved in the outcome. In this way, we aim to avoid involving the Earth International Court of Justice System, thus preserving the interests of the colonies, while still involving the Earth-based origin of the colony in the resolution process. The American Revolution provides a well-known example where the isolation of the thirteen American colonies from their distant rulers in Great Britain led to rising tensions between the interests of the parent nation and the colonies themselves. In this way, we seek to avoid conflict between Earth and any colonies on Mars by preserving solely the interests of the colonies wherever possible.

## 5. Revisiting the Outer Space Treaty

A remaining and important issue in thinking about space colonization is the future of the Outer Space Treaty. Drafted before people had even landed on the Moon, the OST needs to be revisited in the modern era of space exploration, even if the conclusion is to preserve it without modification. The sentiments of the OST are important to uphold, especially as our space presence expands. Respect for the scientific value of planetary bodies, prohibition of military activities off Earth, and a reminder that though we are divided by nationality we all remain equally human. These are significant concepts to remember as our power and our destructive ability grows in space exploration. But before we land on Mars we must reaffirm some increasingly difficult to maintain policies and clarify some vague statements in the OST.

The non-appropriation principle is a particular policy that is difficult to interpret in the context of space settlement. Any country, including all space-faring nations, that signed the OST has agreed not to make any sovereign claim on any part of Mars. This, more than any other part of the treaty, needs to either be discarded or reaffirmed. In the current climate of Mars exploration, the ability of the OST to restrict countries from claiming parts of Mars appears increasingly tenuous. Karl Leib [2] points out an interesting idea which we built upon in our colonization proposal. He questions whether we need to make a sovereign claim on a territory to colonize it. According to the precedent set by the Apollo landings, astronauts representing states may land on, collect materials from, and leave equipment behind on any planet without that suggesting any claim to territory. Could we not simply extend our stay and increase our harvest of materials and remain within the bounds of the treaty? This is an important question to ask, and one that is not clearly answered in the wording of the OST.

Our proposal for allowing exclusive economic zones for each colony fits within the nebulous boundaries imposed by the OST. By claiming only the resources in a radius of land and allowing sovereignty, or control of rules and people, only within the terrestrial space stations themselves, no country will have sovereign reign over Mars territory. This method will have its own challenges as colonies develop, for example, if a second colony decides to settle within another colony's economic zone this may require careful diplomatic resolution to avoid escalation of international tensions on Earth. The examples of international cooperative management shown by the UNCLOS and ATS are encouraging, but these models cannot be precisely applied to Mars in their current form. We must ask ourselves whether allowing some form of national sovereignty would be a more peaceful solution.

The province of mankind principle is another vague policy that needs clarification in order to be properly carried out. Legal uncertainty of any form will deter space development [2]. One way to interpret and implement it is to set up some system where all nations, even ones who are not participating, benefit financially from colonization. This would involve redistributionist policies or some sort of Mars Tax levied on the colonizing countries to be distributed to the non-colonizing countries [4]. Previous experience with some of the treaties mentioned in this paper suggest that redistributionist policies would be unpopular and potentially lead to the rejection of the guidelines outlined here. The Mars Tax might be the more palatable, but probably still unpopular option. Another way to interpret this clause is to claim that even if only some nations are able to participate in Mars colonization, this still represents a step forward in the development of humankind. This interpretation would free colonizing nations from any obligation towards non-colonizing countries. Whatever the resolution, it is important to consider the perspectives of less developed nations as bystanders in an era of space progress [2]. Equatorial developing countries have even attempted to claim the geosynchronous orbit with the Bogota Declaration of 1976 in an attempt to assert their voices in space [27]. Any solution to this problem should keep in mind the interests of these nations, even if they lack the technology to explore space today.

Our proposed pragmatic approach to Mars colonization remains within the legal boundaries of the Outer Space Treaty. The nebulous terminology of the OST allows ample room for interpretation, as well as ample room for conflict to develop. A fundamental reworking of the Treaty may be needed in order to accommodate the activities currently under development concerning Mars. At the very least, a clarification of terminology is needed to avoid future conflict. We suggest a potential clarification to the OST that also remain consistent with our approach of allowing exclusive economic claims on Mars. We recommend that the non-appropriation principle of the OST should be rephrased as: "outer space is not subject to national appropriation by claim of sovereignty, by means of use or occupation *to the exclusion of other nations*, or by any other means;" (emphasis shows added text). The purpose of this distinction is to relax any prohibitions on activities such as mining resources on Mars or asteroids, which would also be consistent with the model of national claims to exclusive economic zones on Mars. An even more radical option may be to remove the phrase 'by means of use or occupation' entirely from the non-appropriation principle, which would fully open up outer space to colonization in a manner similar to historical precedents of colonialism. There are numerous dangers to this route, as well as numerous opportunities for growth. Whatever options decided upon, the OST needs revision in order to remain relevant in the era of modern space exploration.

## 6. Conclusion

In response to rising interest in Mars colonization missions from both national space agencies and private enterprise, we propose a pragmatic solution that allows bounded sovereignty claims outside of preserved planetary parks but remains in the spirit of the Outer Space Treaty. Scientists will establish planetary park locations and regulations. Private and governmental parties may land on and occupy limited plots of Mars' surface, which they will govern according to their national laws. They may conduct science, and may even commence commercial resource exploitation missions. Conflicts will be solved diplomatically, or through use of the temporary tribunal composed of representatives of other colonies on Mars. A Mars Secretariat will act as an administrative body to facilitate communication between the colonies.

Though our proposal is in accord with the OST, we suggest a re-examination of this treaty to clarify uncertain portions of the phrasing of the non-appropriation and province of mankind principles. We also suggest that revisions to the OST relax the restriction on sovereign claims as another valid approach to colonization. We do not deny the importance of any clause of the OST in maintaining peaceful relations in space among nations and upholding the spirit of respect of celestial bodies. We intend our proposal to restrict as much as possible the harmful impacts of human occupation, but also allow the growth of humanity into a multi-planetary species, while minimizing conflict between nations, corporations, and individuals that venture into space.

**Acknowledgements**

This study was conducted during the 2015 Young Scientist Program (YSP) with the Blue Marble Space Institute of Science. The authors thank Sanjoy Som as well as two anonymous reviewers for helpful comments that improved this manuscript. All opinions are those of the authors alone.